\documentstyle{article}
\newcommand{\be}{\begin{equation}} 
\newcommand{\ee}{\end{equation}}
\newcommand{\bea}{\begin{eqnarray}}

\def\abstract#1{\vskip 7mm 
        \begin{center}{\large Abstract}\par \smallskip
                \begin{minipage}[c]{12cm}
                        \small #1
                \end{minipage}
        \end{center}
}
\def\title#1{\begin{center}{\Large\bf #1}\end{center}}
\def\author#1{\vskip 5mm \begin{center}{#1}\end{center}}
\def\address#1{\begin{center}{\it #1}\end{center}}
\makeatletter
\@ifundefined{lesssim}{\def\lesssim{\mathrel{\mathpalette\vereq<}}}{}
\@ifundefined{gtrsim}{}{}
\def\vereq#1#2{\lower3pt\vbox{\baselineskip1.5pt \lineskip1.5pt
\ialign{$\m@th#1\hfill##\hfil$\crcr#2\crcr\sim\crcr}}}
\makeatother

\begin{document}

\title{%
  Spherically Symmetric Self-Similar Solutions and their Astrophysical  and Cosmological Applications}
\author{%
  B.J.Carr\footnote{E-mail:B.J.Carr@qmw.ac.uk}}
\address{%
 Astronomy Unit, Queen Mary and Westfield College,\\
Mile End Road, London E1 4NS, England}
\abstract{
 We discuss spherically symmetric perfect fluid 
solutions of Einstein's equations which have equation of state ($p=\alpha \mu$) and which are self-similar in the sense that all dimensionless variables depend only upon $z\equiv r/t$. For each value of $\alpha$, such solutions are described by two parameters and have now been completely classified. There is a 1-parameter family of solutions asymptotic to the flat Friedmann model at large values of $z$. These represent either black holes or density perturbations which grow as fast as the particle horizon; the underdense solutions may be relevant to the existence of large-scale cosmic voids. 
There is also a 1-parameter family of solutions asymptotic to a self-similar Kantowski-Sachs model at large $z$. These are probably only physically realistic for $-1<\alpha<-1/3$, in which case they may relate to the formation of bubbles in an inflationary universe. There is a 2-parameter family of solutions associated with a self-similar static solution at large $z$. This family contains solutions with naked singularities and this includes the ``critical'' solution discovered in recent collapse calculations for $\alpha < 0.28$. Finally, for $\alpha >1/5$,  there is a family of 
solutions which are asymptotically Minkowski. These asymptote either to infinite $z$, in which case they are described by one parameter, or to a finite value of $z$, in which
case they are described by two parameters and this includes the ``critical'' solution for $\alpha >0.28$. We discuss the stability of spherically symmetric similarity solutions to more general (non-self-similar) spherically symmetric perturbations.
}

\section{Introduction}

Self-similar models have proved very useful in General Relativity because the similarity assumption reduces the complexity of the partial differential equations. Even greater simplification is achieved in the case of spherical symmetry (Cahill \& Taub 1971) since the governing equations then reduce to 
comparatively simple ordinary differential equations. In this case, the solutions can be put into a form in which every dimensionless variable is a function of some dimensionless combination $z$ of the time coordinate t and the comoving radial coordinate r. In the simplest situation, a similarity solution is invariant under the transformation $r \rightarrow ar, t \rightarrow at$ for any constant $a$, so the similarity variable is $z=r/t$. Geometrically this corresponds to the existence of a homothetic vector and is sometimes termed similarity of the ``first'' kind. 

Although we will focus almost exclusively on this kind of similarity here, it should be stressed that there are various ways of generalizing the concept.  For example, there are also similarity solutions of the ``second'' kind, which involve an intrinsic scale. A particularly important application of this is {\it kinematic} self-similarity (Carter \& Henriksen 1991, Coley 1997), in which the similarity variable is of the form $z=r/t^\beta$ for some constant $\beta$.  Ponce de Leon (1993) has also introduced the concept of {\it partial homothety}, in which the Killing condition is only satisfied with spatial hypersurfaces. In this case, the similarity variable has the form $z=r/F(t)$ for some general function $F$ and it reduces to self-similarity of the first kind only if $F$ is linear in t. However, this approach is not covariant.  There are also solutions which possess {\it discrete} self-similarity, in the sense that dimensionless variables repeat themselves on some spacetime scale $\Delta$, i.e. they are invariant under the transformation $r \rightarrow e^{-n\Delta}r, t \rightarrow e^{-n\Delta}t$ for any integer $n$. One recovers the {\it continuous} type of self-similarity discussed above in the limit $\Delta \rightarrow 0$. Although discrete self-similarity is much harder to deal with mathematically, Choptuik (1997) has emphasized its crucial role in the context of critical phenomena. 

What makes similarity solutions of more than mathematical interest is the fact that they are often relevant to the real world dynamics (Barenblatt \& Zeldovich 1972). This is because, in a variety of astrophysical 
and cosmological situations, solutions may naturally evolve to self-similar form even if they start out more complicated. For example, an explosion in a homogeneous background produces
fluctuations which may be very complicated initially but which
tend to be described more and more closely by a spherically
symmetric similarity solution as time evolves (Sedov 1967). This
applies even if the explosion occurs in an expanding cosmological 
background (Schwartz et al. 1975, Ikeuchi et al. 1983). The evolution of cosmic ``voids'' is also described by a similarity solution at late times (Hoffman et al. 1983, Hausman et al. 1983, Maeda \& Sato 1984, Bertschinger 1985).  In the hierarchically clustering scenario gravitationally bound clouds tend to form self-similar cosmic structures (Gunn \& Gott 1972, Gunn 1977, Fillmore \& Goldreich 1984, Bertschinger \& Watts 1984) and a cloud collapsing from an initially uniform static configuration may also evolve towards self-similarity (Larson 1969, Penston 1969). Recently it has become clear that spherically symmetric self-similar solutions also play a crucial
role in the context of the ``critical" phenomena (Choptuik 1993, Evans and Coleman 1994, Gundlach 1995, Koike et al. 1995, Maison 1996, Carr et al. 1999).

These examples suggest that self-similar solutions may play the same sort of role in describing the asymptotic properties of spherically symmetric models as they do in the {\it spatially homogeneous} context (Hsu \& Wainwright 1986).  In the cosmological context, these considerations have led to the ``similarity hypothesis'' (Carr 1993), which proposes that - under certain circumstances (e.g. non-zero pressure, non-linearity, shell-crossing etc.) - cosmological solutions may naturally evolve to a self-similar form.  One might also extend this hypothesis to the more general spherically symmetric context. 

The possibility that self-similar models may be singled out in this way from more general spherically symmetric solutions means that it is essential to understand the full family of similarity solutions. This paper will focus exclusively on the situation in which the source of the gravitational field is a perfect fluid with equation of state of the form $p = \alpha \mu$ (where $\alpha$ is a constant). Indeed, Cahill \& Taub (1971) have shown that this is the only equation of state compatible with the similarity assumption. In this case, it is well known that, for a given value of $\alpha$, spherically symmetric similarity solutions are described by two parameters. Such solutions have now been classified completely (Carr \& Coley 1999a; CC). We describe the characteristics of these solutions in Section 2 and discuss some of their cosmological and astrophysical applications in Section 3. In completing this introduction, we also summarize the main results here.

CC find that there are four classes of solutions and each of these relate to a specific application. The first class consists of the 1-parameter family of solutions asymptotic to the exact $k =0$ Friedmann model at large $z$ and these are obviously relevant in the cosmological context (Section 2.3). Some of
them contain black holes which grow at the same rate as the particle horizon (Carr \& Hawking 1974, Bicknell \& Henriksen 1978a). Others represent density
perturbations in a Friedmann background which always maintain the same form relative to the particle horizon (Carr \& Yahil 1990). The latter all contain a sonic point and the requirement that they be regular there severely restricts their form (Bogoyavlenski 1985). While there is a continuum of regular underdense solutions, regular overdense solutions only occur in narrow bands. The underdense solutions may be relevant to the existence of large-scale cosmic voids (Section 3.3). The overdense solutions do not seem to describe features in the present Universe but they may relate to the formation of  overdense blobs at an early phase transition.

The second class of models are asymptotic to a self-similar
member of the Kantowski-Sachs (KS) family (Section 2.4).
For each non-zero $\alpha$ there is a unique self-similar KS solution and there also exists a $1$-parameter family of solutions asymptotic to this at both large and small values of $z$ (Carr \& Koutras 1992). Solutions with $-1/3 < \alpha <1$ are probably unphysical because they are tachyonic and the mass is negative. 
Solutions with $-1< \alpha < -1/3$ avoid these problems and are
therefore more interesting. Although this equation of state violates the strong energy 
condition, it could could well arise in the early Universe. Indeed, such models may be related to the growth of bubbles formed at a phase transition in an inflationary model (Wesson 1986). Generalized negative-pressure KS solutions, in which the similarity is of the second kind, have been studied by Ponce de Leon (1988) and Wesson (1989); in these the equation of state deviates from $p=\alpha \mu$ and the fluid can be interpreted as a mixture of false vacuum and dust. 

The third class of models are associated with self-similar static models. There is just one static self-similar solution for each positive value of $\alpha$ (Misner \& Zapolsky 1964) and 
there is also a 1-parameter family of solutions asymptotic to this. However, there is a 2-parameter family of solutions which 
are asymptotically ``quasi-static'' in the 
sense that they have an isothermal density profile at large values of $z$ (Section 2.5). Since the asymptotically Friedmann and asymptotically KS solutions are described by only one parameter, this third class of models play a crucial role in
understanding the full family of similarity solutions. Some asymptotically quasi-static solutions have been studied by Foglizzo \& Henriksen (1993) and
a more extensive analysis is being carried out by Carr et al. (1999). Such solutions are of particular interest because they are are associated with the formation of naked singularities (Section 3.1) and, for $\alpha <0.28$, the occurrence of critical phenomena (Section 3.2).

The fourth class of solutions, which only exist for $\alpha >1/5$, are asymptotically Minkowski and have not been previously analysed at all. They were originally 
found numerically by Goliath et al. (1998b) and this led CC to ``predict" them analytically. There are actually two such families and they are described in more
detail  elsewhere (Carr et al. 1999). Members of the first family are described by one parameter and are asymptotically Minkowski as $|z|\rightarrow \infty$; members
of the second family are described by two parameters and are asymptotically Minkowski as $z$ tends to some finite value (though this corresponds to an infinite
physical distance unless $\alpha=1$). As with the asymptotically Friedmann and asymptotically quasi-static solutions, these may
be either supersonic everywhere (in which case they contain a black hole or naked singularity) or attached to $z=0$ via a sonic point (in which case
they are asymptotically Friedmann or exactly static at small $|z|$). The transonic ones are associated with critical phenomena for $\alpha > 0.28$ (Carr et al. 1999).

\section{Spherically Symmetric Similarity Solutions}

One can analyse self-similar spherically symmetric models using three possible approaches. The first one uses ``comoving'' coordinates and was pioneered by Cahill \& Taub (1973) and then followed by Carr and Henriksen and coworkers. The second approach, followed by  
Bogoyavlenski (1977) and coworkers, uses ``homothetic'' coordinates in which the homothetic Killing vector is along either the time or space axis. In this case, the equations can be reduced to those of a dynamical system and one can exploit results derived from the study of hypersurface homogeneous models (cf. Goliath et al. 1998a, 1998b). A third approach uses
``Schwarzschild'' coordinates and was included in the analysis of Ori \& Piran (1990) and Maison (1996). This is useful if one is matching the solution to an asymptotically flat spacetime and wishes to find its global structure. All of these approaches are complementary and which is most suitable depends on what type of problem one is studying. In this paper we wish to
emphasize the form of the solutions explicitly and it is most convenient to use the first approach.
 
\subsection{General features of similarity solutions}

In the spherically symmetric situation one can introduce a time coordinate t such that surfaces of constant t are orthogonal
to fluid flow lines and comoving coordinates $(r, \theta, \phi)$ 
which are constant along each flow line. The metric can be written in the form
\be
\label{lelement}
        ds^{2}=e^{2\nu}\,dt^{2}-e^{2\lambda}\,dr^{2}-R^{2}\,d\Omega^{2},\;\;\;
        d\Omega ^2 \equiv d\theta^{2}+\sin^{2}\theta \,d\phi^{2} 
\ee
where $\nu$, $\lambda$ and $R$ are functions of $r$ and $t$. The equations have a first integral
\be
\label{firstint}
   m(r,t)=\mbox{$\frac{1}{2}$} R\left[ 1+e^{-2\nu}
   \left(\frac{\partial R}{\partial t}
   \right)^{2} - e^{-2\lambda}\left(\frac{\partial R}
   {\partial r}\right)^{2}\right]
\ee
and this can usually be interpreted as the mass within comoving radius $r$ at time $t$: 
\be
\label{massfunct}
   m(r,t)=4\pi\int_{0}^{r}\mu R^{2}\frac{\partial R}{\partial r'}\,dr'.
\ee
There is also a dimensionless quantity $E(r,t)$ which represents the total energy per unit mass for the shell with comoving
coordinate $r$. Unless $p=0$, both these quantities decrease with increasing $t$ because of the work done by the pressure.

Spherically symmetric homothetic solutions were first
investigated by Cahill \& Taub (1971), who showed that by a suitable coordinate transformation they can be put into a form in which all dimensionless quantities such as $\nu$, $\lambda$, $E$ and
\be
   S\equiv\frac{R}{r},\;\;\;\;M\equiv\frac{m}{R},\;\;\;\;
   P\equiv pR^{2},\;\;\;\;W\equiv\mu R^{2}
\ee
are functions only of the dimensionless variable $z\equiv r/t$.  Values of $z$ for which $M=1/2$ correspond to a black hole or cosmological apparent horizon
since the congruence of outgoing null geodesics have zero divergence. Another important quantity is the function
\be
\label{velocity}
        V(z)=e^{\lambda -\nu}z
\ee
which represents the velocity of the spheres of 
constant $z$ relative to the fluid. These spheres contract relative to the fluid for $z<0$ and expand for $z>0$. Special significance is attached to values of $z$ for which $|V|=\sqrt{\alpha}$ and $|V|=1$. The first corresponds to a sonic point (where the pressure
and density gradients are not uniquely determined), the second to a Cauchy horizon (either a black hole event horizon
or a cosmological particle horizon).

The only barotropic equation of state compatible with the similarity ansatz is
one of the form $p=\alpha \mu$, where causality requires $-1\leq\alpha\leq 1$. It is convenient to introduce a dimensionless function $x(z)$ defined by
\be
\label{xdef}
        x(z)\equiv (4\pi \mu r^{2})^{-\alpha/(1+\alpha)}.
\ee
The conservation equations $T^{\mu \nu}_{\;\;\;\; ;\nu}=0$ can then be 
integrated to give
\be
e^{\nu}=\beta x z^{2\alpha/(1+\alpha)},\;\;\;
e^{-\lambda}=\gamma x^{-1/\alpha} S^{2}
\ee
where $\beta$ and $\gamma$ are integration constants. The remaining field equations reduce to a set of ordinary differential equations in $x$ and 
$S$ (given explicitly in CC). These specify integral curves in the
$3$-dimensional $(x, S, \dot{S})$ space (where a dot denotes $zd/dz$). For a given equation of state parameter $\alpha$, there is therefore a
$2$-parameter family of spherically symmetric similarity solutions.

In $(x, S, \dot{S})$ space the sonic condition $|V|= \sqrt{\alpha}$ specifies a $2$-dimensional surface. Where a curve intersects this surface, the equations do not uniquely
determine $\dot{x}$ (i.e. the pressure gradient), so there can be a number of different solutions passing
through the same point. However, only integral curves which pass through a line $Q$ on the sonic surface are ``regular'' in the sense that $\dot{x}$ is finite, so that they can be extended beyond there. The equations permit just two values of $\dot{x}$ at each point of $Q$ and these values will be real only on a restricted part of it. On this part, there will be two corresponding values of $\dot{V}$, at least one of which will 
be positive.  If both values of $\dot{V}$ are positive, the smaller one is 
associated with a $1$-parameter family of solutions, while the larger one
is associated with an isolated solution.  If one of the values of $\dot{V}$ is
negative, both values are associated with isolated solutions. This behaviour has been analysed in detail by Bogoyavlenski (1977), Bicknell \& Henriksen (1978a), Carr \& Yahil (1990) and Ori \& Piran (1990).

On each side of the sonic point, $\dot{x}$ may have either of the two 
values.  If one chooses different values for $\dot{x}$, there will be a discontinuity in the pressure gradient.  If one chooses the same value,
there may still be a discontinuity in the higher derivatives of $x$, in which case the solutions are described as ``weakly discontinuous''.  Only 
the isolated solution and a single member of $1$-parameter family of solutions  are {\it analytic}. This contrasts with the case of a shock where $x$
is itself discontinuous (Cahill \& Taub 1971, Bogoyavlenski 1985,
Anile et al. 1987).  One can show that the part of $Q$ for which there is a $1$-parameter family
of solutions corresponds to two ranges of values for $z$. One range $(z_1 < z < z_2)$ lies to the left of the Friedmann sonic point $z_F$ and includes the static sonic point $z_S$. The other range $(z > z_3)$ includes the Friedmann sonic point. The ranges for $\alpha =1/3$ are indicated in Figure (1); in this case, $z_2=z_S$ and $z_3 =z_F$. 

Carr \& Coley (1999a; CC) have classified the $p=\alpha \mu$ spherically symmetric similarity solutions completely. The key steps in their analysis are: (1) a complete analysis of the dust solutions, since this
provides a qualitative understanding of the
solutions with pressure in the supersonic regime; (2) an elucidation of the link between the $z>0$ solutions and the $z<0$ solutions; (3) a proof that, at large and small values of $|z|$, all similarity solutions must have an asymptotic form in which $x$ and $S$ have a power-law dependence on $z$; (4) a demonstration that there are only three such power-law solutions; and  (5) a proof that solutions whose asymptotic behaviour is associated with a {\it finite} value of $z$ have a power-law dependence on $lnz$. We first discuss the power-law solutions and then the asymptotic ones. We will assume $z>0$ in the first case but we will need to consider both signs of $z$ in the second. The equations can be extended to the $z<0$ regime by replacing $z$ by $|z|$ and reversing the sign of $V$. Except in the case of the KS or asymptotically KS solutions, we also assume $\alpha >0$.

\begin{figure}
\vspace*{4.5in}
\caption{The form of $V(z)$ for the exact Friedmann, static and
Kantowski-Sachs solutions in the $\alpha=1/3$ case. Also shown  are the curves corresponding to $M=0$ (solid), the sonic lines
$|V|=1/\sqrt{3}$ (broken) and the range of values of $z$
in which there is a 1-parameter family of regular solutions at the sonic point.}
\end{figure}

\subsection{Power-law solutions}

\indent $\bullet$ The $k=0$ Friedmann solution. For this one can choose $\beta$ and $\gamma$ in eqn (7) such that
\be
        x=z^{-2\alpha/(1+\alpha)},\;\;\;\;
        S=z^{-2/[3(1+\alpha)]}
\ee
and then
\be
\label{friedvel}
        \mu = \frac{1}{4\pi t^2},\;\;\;
        V=\left(\frac{1+3\alpha}{\sqrt{6}}\right)
        z^{(1+3\alpha)/[3(1+\alpha)]}.
\ee
One can put the metric in a more familiar form by making the coordinate transformation
\be
   {\hat{t}}=\beta t,\;\;\;\;
   {\hat{r}}=\beta^{-2/[3(1+\alpha)]}r^{(1+3\alpha)/[3(1+\alpha)]}
\ee
which gives
\be
   ds^{2}=d{\hat{t}}\;^{2}-{\hat{t}}^{4/[3(1+\alpha)]}[d{\hat{r}}^{2}+
   {\hat{r}}^{2}d\Omega^{2}].
\ee
Only the $k=0$ Friedmann model is self-similar since the $k=\pm 1$ models contain an intrinsic scale.

$\bullet$ A self-similar Kantowski-Sachs model. For each $\alpha$ there is a unique self-similar KS solution and this can be put in the form
\be
S=S_0z^{-1}, \;\;\; x=x_0z^{-2\alpha/(1+\alpha)}
\ee
where $x_0$ and $S_0$ are determined by $\alpha$. It is convenient to take $\beta$ and $\gamma$ to have the same values as for the Friedmann solution if
$\alpha < 0$ and i times those values if $\alpha > 0$. In this case, one has
\be
   \mu t^{2} =  \frac{1}{4\pi}\; \left( \frac{1}{3|\alpha|}\right)^{(1+\alpha)/(\alpha-1)},\;\;\;
   V = -\frac{(1-\alpha)(1+3\alpha)^2}{2 \sqrt{6}\alpha} \left( \frac{1}{3|\alpha|}\right)^{-2 \alpha/(1-\alpha)} z^{(1+ 3 \alpha)/(1+\alpha)}
\ee
$V$ is negative for $z>0$, corresponding to tachyonic solutions, if $0< \alpha <1$; this means that the $t $ coordinate is spacelike and the $r$ coordinate is timelike. For $-1/3<\alpha<0$, $V$ is positive but $S$ is imaginary. The mass $m$, which depends only on $t$ in the KS case, is also negative for $-1/3<\alpha<0$. [The KS solution has no origin, so eqn (3) is inapplicable.] Therefore probably only solutions with $\alpha < -1/3$ are physical. One can put the metric in a more familiar form by making the coordinate transformation
\be
   \hat{t}= \beta x_0 t,\;\;\; \hat{r}= \gamma^{-1}
   (\beta x_0)^{2\alpha/(1 + \alpha)} x_0\,^{1/\alpha} S^{-2}_0\,
   r\,^{(1 + 3 \alpha)/(1 + \alpha)}.
\ee
This gives 
\be
  ds^2 = d\hat{t}^2 - \hat{t} ^{-4 \alpha/(1 + \alpha)} d \hat{r}^2
  - (S_c/\beta x_c)^2\hat{t}^2 d \Omega^2 
\ee
although it should be noted that the new coordinates may not be real for $\alpha > -1/3$. 

$\bullet$ A self-similar static solution. In this case 
\be
x=x_0,\;\;\;S=S_0
\ee
where the constants $x_0$ and $S_0$ are determined uniquely by $\alpha$, so there is just one static solution for each equation of state. One also has  
\be
   \mu = x_0^{-(1+\alpha)/\alpha}(4\pi r^2)^{-1},\;\;\;
    V = \sqrt{3\alpha}\; x_0^{(1-\alpha)/2\alpha}z^{(1- \alpha)/(1+\alpha)},\;\;\;
\ee
where the density profile has the usual isothermal form. Note that there is a naked singularity at the origin. By introducing the variables
\be
\hat{r} = rS_0,\;\;\;\hat{t} = \left(\frac{1+\alpha}{1-\alpha}\right) \beta x_0 S_0^{-2\alpha/(1+\alpha)},
\ee 
the metric can be put in an explicitly static form
\be
ds^2 = \hat{r}^{4\alpha/(1+\alpha)}d\hat{t}^2 - \gamma^{-2}x_0^{2/\alpha}S_0^{-6}d\hat{r}^2 - \hat{r}^2 d\Omega ^2.
\ee
The existence of the self-similar static solution has been noted by several authors (Misner \& Zapolsky 1964, Henriksen \& Wesson 1978a \& 1978b, Carr \& Yahil 1990). There is an interesting connection between the static and KS solutions:  if one interchanges the $r$ and $t$ coordinates in the static metric and also changes the equation of state parameter to
\be
   \alpha' = -\frac{\alpha}{1+2\alpha},
\ee
one obtains the KS metric. For a static solution with a normal equation of state $(1>\alpha>0)$, $\alpha'$ must lie in the range $-1/3$ to $0$, so negative pressure (negative mass) KS solutions can also be interpreted as positive pressure (positive mass) static solutions.

The forms of $V(z)$ for the Friedmann, KS and static solutions in the $\alpha =1/3$ case are shown in Figure (1). To study the family of similarity solutions asymptotic to these or the associated $z<0$ solutions at large and small $|z|$, one introduces functions $A(z)$ and $B(z)$ defined by
\be
   x \equiv x_i e^A,\;\;\;
   S\equiv x_i e^B 
\ee
where $x_i$ is given by eqn (8) in the Friedmann case, eqn (12) in the KS case and eqn (16) in the static case. The ordinary differential equations for $x$ and $S$ then become ordinary differential equations for $A$ and $B$.
The solutions in each family can be specified by the values of $A$ and $B$ as $|z| \rightarrow \infty$ and $|z| \rightarrow 0$, although these values may not be independent. The form of the complete family of solutions in the $\alpha =1/3$ case is
summarized in Figure 6.

\subsection{Asymptotically Friedmann solutions} 

In the supersonic (large $z$) regime, one linearizes the equations in $A$, $\dot{A}$ and $\dot{B}$ and obtains the 1st order solution
\be
   A = - \frac{\alpha (1 + 3 \alpha)}{(1+ \alpha)} kz ^{-2(1+3   \alpha)/3(1+\alpha)},\;\;\;
   B = B_\infty -kz^{-2(1+3\alpha)/3(1+\alpha)}
\ee
where $B_\infty$ and $k$ are integration constants related by
\be
   k = \frac{3(1+\alpha)(e^{-2B_\infty} -e^{4B_\infty})}{2(1+3 \alpha)
  (5+3  \alpha)} .
\ee
Thus there is a $1$-parameter family of asymptotically Friedmann solutions and one can take this parameter to be $B_\infty$. Eqns (6), (21) and (23) show that solutions are overdense relative to the Friedmann solution for $B_\infty < 0$ and underdense for $B_\infty > 0$. 

If $B_\infty$ is sufficiently negative, $V$ reaches a minimum value and then rises again to infinity as $z$ decreases. Such solutions contain black holes and were originally studied because there was
interest in whether black holes could grow at the same rate as the particle horizon. Carr \& Hawking (1974) showed that such solutions exist for radiation ($\alpha=1/3$) and dust ($\alpha=0$) 
but only if the universe is asymptotically rather than exactly 
Friedmann ($B_\infty \neq 0$), i.e. one cannot attach a black hole to an exact Friedmann model via a sound-wave. This means that  black holes formed through purely local processes
in the early Universe cannot grow as fast as the particle horizon.
Carr (1976) and Bicknell \& Henriksen (1978a) then extended this result to a general $0<\alpha<1$ fluid. Lin et al. (1978) claimed that a black hole
similarity solution {\it can} be attached to an exact Friedmann model for the special case of a stiff fluid ($\alpha=1$) but Bicknell \& Henriksen (1978b) showed that this requires the inflowing material to turn into a radiation fluid at the event horizon. In fact, 
for fixed $\alpha$, it is now known that {\it all} subsonic solutions which can be attached to an exact Friedmann model via a sound wave are non-physical:
as one goes inward from the sound wave they either 
enter a negative mass regime or reach another sonic point at which
the pressure diverges (Bicknell \& Henriksen 1978a). 

It is likely that 
asymptotically Friedmann solutions which contain black holes are supersonic everywhere (in the sense that V never falls below $1/\sqrt{\alpha}$), although this has not been rigorously proved.  However, all the solutions with $B_\infty$ exceeding some critical negative value $B_\infty^{crit}$ reach the sonic surface and those for which the value of $z$ at the sonic point lies within the ranges $z_1 < z < z_2$ and $z > z_3$ [indicated in Figure (1) for $\alpha =1/3$] may be attached to the origin by subsonic solutions. The latter are also described by a single parameter: $A$ and $B$ tend to related constants as $z \rightarrow 0$, so we can take the parameter to be $A_0$, the value of $A$ at $z=0$. This is a measure of the overdensity at the origin since eqns (6) and (21) imply
\be
A_0 = \left[\frac{\alpha}{1 + \alpha}  \right] \log \left[  \frac{\mu_F(0)}{\mu(0)}\right], 
\ee
where $\mu_F$ is the density in the Friedmann solution. Thus solutions are overdense relative to the Friedmann solution for $A_0<0$ and underdense for $A_0>0$. These transonic solutions represent density fluctuations in a flat Friedmann background which grow at the same rate as the particle horizon (Carr \& Yahil 1990). While there is a continuum of regular underdense solutions, regular overdense solutions only occur in narrow bands (with just one solution per band being analytic). The overdense solutions exhibit oscillations inside the sonic point, with the number of oscillations labelling the band. The existence of these bands was first pointed out by Bogovalenski (1985) and also studied by Ori \& Piran (1990). The  band structure arises even in the Newtonian situation (Whitworth \& Summers 1985). The higher bands are all nearly static near the sonic point ($z_1 = z_S$ if $\alpha =1/3$), although they
deviate from the static solution as one goes towards the origin. The form of the solutions near the sonic point is qualitatively similar for all values of $\alpha$. It is illustrated in Figure (2) for $\alpha =1/3$, the case most likely to pertain in the early Universe, although this shows only the first overdense band explicitly.

\begin{figure}
\vspace*{4.0in}
\caption{The form of the velocity function $V(z)$ near the sonic
line for the asymptotically Friedmann solutions with $\alpha=1/3$. In the supersonic regime, they are described by 
the parameter $B_\infty$. The spacing is 0.05 for $B_\infty>0$ and 0.01 for $B_\infty<0$, except in the range between 0.13 and 0.15, where it is
0.02. In the subsonic regime, solutions are described by 
the parameter $A_0$ with spacing 0.1. The solutions which are regular on both sides of the sonic point are shown by solid curves; otherwise they are shown dashed, with long dashes for solutions containing black holes.}
\end{figure}

The overall form of $S(z)$ and $V(z)$
in the $\alpha =1/3$ asymptotically Friedmann solutions is indicated in Figure (3). For comparison  with the other similarity solutions, the curves are here parametrized by the ``asymptotic energy'' parameter ($E$), which is related to $B_\infty$ by
\be
E = \frac{1}{2} (e^{6B_\infty} -1).
\ee
The $z>0$ solutions correspond to inhomogeneous models which start from an initial Big Bang singularity at $z=\infty$ ($t=0$) and then, as $z$ decreases, either expand to infinity for $E >E_{crit}$ or recollapse for $E<E_{crit}$; here $E_{crit}$ is the (negative) value associated with $B_\infty^{crit}$. The recollapsing solutions contain a black hole event horizon and a cosmological particle horizon for values of $E$ exceeding another negative value $E_*$ such that $V_{min}$ is below 1. In this case, the mass of the black hole is
\be
m_{BH}=(MSz)_{BH}\,t,
\ee
so it forms with zero mass at $t=0$ and then grows as $t$. We have seen that there are overdense oscillating solutions without black holes providing $E $ lies in narrow bands between $E_{crit}$ and $0$
but Figure (3) does not show these oscillations explicitly. The  analysis is trivially extended to the $z<0$ regime. The $z<0$ solutions are merely the time-reverse of the $z>0$ ones: $E >E_{crit}$ models collapse from an infinitely dispersed initial state to a Big Crunch singularity as $z$ decreases from 0 to $-\infty$ (i.e. as $t$ increases from $-\infty$ to 0), while $E <E_{crit}$ models emerge from a white hole and are never infinitely dispersed.

\begin{figure}
\vspace*{7.5in}
\caption{This shows the form of the scale factor $S(z)$ and the velocity function $V(z)$ for the asymptotically Friedmann solutions 
with different values of $E$. The $z>0$ ($z<0$) solutions contain black (white) holes for $E* < E < E_{crit}$, so that the minimum of $|V|$ lies between 1 and $\sqrt{\alpha}$. The solutions with $E > E_{crit}$ contain a sonic point (bold). }
\end{figure}

\subsection{Asymptotically Kantowski-Sachs solutions}

The asymptotically KS models were studied by Carr \& Koutras (1992). At large values of $|V|$, linearizing the equations in $A$, $\dot{A}$ and $\dot{B}$ gives the 1st order solution 
\be
A = A_\infty z^{-p_{1,2}},\;\;\; B = A_\infty \left[\frac{1}{2 \alpha} - \left(\frac{1 - \alpha}{1 + \alpha}\right) \frac{1}{p_i}\right]z^{-p_{1,2}}, 
\ee
where $A_\infty$ is an integration constant and
\be
p_{1,2} = \frac{-1 + \alpha \pm \sqrt{(1 - \alpha)(24\alpha^2 + 7 \alpha +1)}}{2 (1+\alpha)}.
\ee
There is thus a 1-parameter famly of solutions. For $\alpha > 0$, the KS solution has $|V| \rightarrow \infty$ as $z \rightarrow \infty$, so we must choose the
positive root $p_1$. For $-1< \alpha < - 1/3$, the KS solution has
$|V| \rightarrow \infty$ as $z \rightarrow 0$, so we must choose the negative root $p_2$.  For $-1/3 < \alpha < 0$, one again has  $|V| \rightarrow \infty$ as $z \rightarrow \infty$ but both $p_1$ and $p_2$
are negative, so there are no solutions. At small values of $|V|$, $A$ and $B$ must be constant but their values are related, so there is again a $1$-parameter family of solutions. The parameter may be taken to be $A_0$ and is a measure of the underdensity or overdensity relative to the KS solution. 

Solutions with $\alpha >0$, like the KS similarity solution itself, are probably unphysical since the mass is negative. It also turns out that there are only isolated solutions at a sonic point in this case, so solutions which hit the sonic surface are unlikely to be regular there. However, these solutions are still of mathematical
interest, since they fill in the negative $V$ quadrant in Figure (6).
The solutions with $-1 < \alpha < -1/3$ may be more physical since this equation of state could arise in the early Universe due to inflation or particle production. Such models may be related to the growth of $p>0$ bubbles formed at a phase transition in a $p<0$ cosmological background (Wesson 1986). Note that the asymptotic energy in all these solutions is $E= -1/2$, which from eqn (25) corresponds to the
limit of the asymptotically Friedmann solutions as
$B_\infty \rightarrow -\infty$. 

A rather peculiar feature of the asymptotically KS solutions is that the mass
can go negative. Indeed this is a general feature of similarity solutions which can occur even for $V>0$. This may seem unphysical but - in the context of the Big Bang model - Miller (1976) has given a possible interpretation of this in terms of ``lagging'' cores. She gives an explicit example of a self-similar solution for which the mass goes negative in the stiff ($\alpha =1$) case. In the $\alpha =1/3$ case (but only this case), one can show that there is a 
well-defined curve in the $V(z)$ diagrams where $M=0$:
\be
V^3 = -\sqrt{3/2} \, z^{3/2}(V^2- 1/9).
\ee
This curve is shown in
Figure (1) and has two parts. The upper part (with $V > 0$) is relevant for asymptotically Friedmann solutions, while the lower part (with $V < 0$) is 
relevant for asymptotically KS solution. $M$ is negative in between the
two parts and this region includes KS itself (as expected). 

\subsection{Asymptotically quasi-static solutions}

For $0<\alpha<1$,  the 1st order solution as $V \rightarrow \infty$ (i.e. as $z \rightarrow \infty$), is 
\be
A = A_{\infty} + Cz^{-(1-\alpha)/(1+\alpha)},\;\;\;
B = B_{\infty} + \left(\frac{C}{2\alpha}\right) z^{-(1-\alpha)/(1+\alpha)}
\ee
where $C$ is determined by the arbitrary integration constants $A_{\infty}$ and $B_{\infty}$. These constants also determine the asymptotic energy parameter
\be
E_\infty = \frac{(1+\alpha)^2}{2(1+6\alpha + \alpha^2)}e^{6B_\infty - 2A_\infty /\alpha} - \frac{1}{2}. 
\ee
This shows that asymptotically static solutions are described by {\it two} parameters at infinity for a given equation of state. They are therefore of particular interest since they represent the most general solution. 

Describing these solutions as ``asymptotically static'' is rather misleading because one can show that the velocity of the fluid relative to the constant $R$ surfaces is
\be
V_R \approx - \frac{(1-\alpha)}{2\alpha(1+\alpha)} C,
\ee
at large $z$, so only the 1-parameter family of solutions with $C=0$ are strictly static at infinity. This agrees with the description of Foglizzo \& Henriksen (1993), who term $C=0$ solutions ``symmetric''. (The exact static solution has $A_{\infty}=B_{\infty}=C=0$ and is just one member of this family.) CC describe the more general solutions with $V_R\neq 0$ as asymptotically ``quasi-static'' since they still have $\dot{S}$ and $\dot{x}$ tending to 0 at infinity and one can show that they also have an isothermal density profile there. 

These solutions can be straightforwardly extended to the $z<0$ regime; indeed they {\it need} to be since the complete asymptotically quasi-static solutions necessarily span both positive and negative values of $z$. This is illustrated by the form of $S(z)$, given in Figure (4). If $E$ exceeds some critical negative value $E_{crit}$, there are monotonically expanding and collapsing solutions, each of which start in the $z<0$ region and then cross to $z>0$. (This contrasts with the asymptotically Friedmann case, in which the expanding and collapsing solutions are confined to 
$z>0$ and $z<0$, respectively.) The expanding solutions can be interpreted as inhomogeneous Big Bang models but with the initial singularity at $z=-1/D$ rather than $z=\infty$ (i.e. for each fluid element the singularity occurs before $t=0$). Here $D$ is a positive constant which depends implicitly on $A_{\infty}$ and $B_{\infty}$, although it cannot be expressed in terms of them explicitly. The collapsing solutions are the time reverse of the expanding ones and describe the collapse of an inhomogeneous gas cloud to a Big Crunch singularity at $z=+1/D$ (i.e. after $t=0$). For $E<E_{crit}$, $S$ never reaches infinity, so the models either expand from a Big Bang singularity at $z=-1/D$ and then recollapse to another singularity at $z_S$ or they expand from a white hole singularity at $-z_S$ and then recollapse to a Big Crunch singularity at $z=1/D$. There is a minimum value of $E$, associated with the symmetric solution for which $z_S=1/D$, and this is denoted by $E_{sym}$ in Figure (4). 

The form of $V(z)$ in these solutions is also indicated in Figure (4). The monotonically collapsing solutions start with $V=0$ at $z=0$ and then, as $z$ decreases, pass first through a sonic point (where $V=-\sqrt{\alpha}$) and then a Cauchy horizon (where $V=-1$), before tending to the quasi-static form at $z=-\infty$. They then jump to $z=+\infty$ and enter the $z>0$ regime. As $z$ further decreases, $V$ first reaches a minimum and then diverges to infinity when it encounters the singularity at $z=1/D$. The minimum of $V$ will be below 1 if $E$ exceeds some value $E_+(D)$ and, in this case, one necessarily has a naked singularity. The singularity forms with zero mass at $t=0$ but its mass $m_S$ then grows as $t $ since
\be
m_S=(MSz)_S\;t.
\ee
As pointed out by Ori \& Piran (1990), such solutions have an analogue in Newtonian theory (Larson 1969, Penston 1969). Solutions in which the minumum of $V$ falls below $\sqrt{\alpha}$ can probably be excluded since there would need to be two sonic points and the solution is unlikely to be regular at both of them. The monotonically
expanding solutions are just the time reverse of the collapsing ones. The expanding-recollapsing solutions with $E<E_{crit}$ have no sonic points. The ones which expand from $z=-1/D$ (where $V=-\infty$) and collapse to $z=z_S$ (where $V=+\infty$) will have a black hole event horizon and cosmological particle horizon for  $E_*<E<E_{crit}$, corresponding to the minimum of $V$ being less than 1.

It should be noted that the introduction of the ``second" parameter $D$ has relatively little effect on the form of the solutions in the subsonic regime. Indeed one can show that all solutions must be either {\it exactly} static or asymptotically Friedmann at small $|z|$, so one can still use the analysis in Section 2.3 here.

\begin{figure}
\vspace*{8.0in}
\caption{This shows the form of the scale factor $S(z)$ and the velocity function $V(z)$ for the asymptotically quasi-static solutions
with different values of $E$ and $D$. The solid curves show solutions which expand from an initial
singularity at $z=-1/D$ and then either expand forever if $E>E_{crit}(D)$, in which case they pass through a sonic point (bold), or 
recollapse to another singularity if $E<E_{crit}(D)$. There is an event horizon and particle horizon for
$E_*(D)<E<E_{crit}(D)$. The last recollapsing solution is the symmetric one for which $E=E_{sym}$. The broken curves are the time reverse of the solid ones.}
\end{figure}

\subsection{ Asymptotically Minkowski Solutions}

For $\alpha >1/5$ there are also two families of asymptotically Minkowski self-similar 
solutions which were only discovered very recently, one associated with infinite
$z$ and the other with a finite value. These solutions are discussed in more detail by 
Carr et al. (1999) and turn out to play
an important role in the context of critical phenomena. The forms of $S(z)$ and $V(z)$ are illustrated in Figure (5).

The first family
is associated with the limit $z\rightarrow \infty$ and is characterized by the fact that 
$V$ tends to a finite value
\be
V_* = \frac{\alpha(1+\alpha) + \sqrt{\alpha(\alpha^3 - \alpha^2 + 3\alpha +1)}}{1 - \alpha}.
\ee
This exceeds $1$ if $\alpha >1/5$; solutions with $\alpha <1/5$ are unphysical in the sense that the mass
is negative. In addition, 
\be
x=x_{o} z^{a},\;\;\;S=S_{o} z^{b}
\ee
where $a$ and $b$ have a complicated dependence on $\alpha$.
Both exponents are positive for $\alpha >1/5$, so the scale factor tends to infinity, while the density tends to zero as $z\rightarrow \infty$. The mass function $M$ also tends to zero,
so these solutions are asymptotically flat, although it requires a complicated coordinate
transformation to demonstrate this explicitly. Eqns (34) and (35) impose a relationship between $x_0$ and $S_0$, so these solutions are described by just one independent parameter.

The second family of asymptotically flat solutions has $V\rightarrow 1$ at some finite value $z_*$. In this case, it can be shown that
\be
S =S_0 L^{(1-\alpha)/(1-5\alpha)},\;\;\;x = x_0 L^{2\alpha/(1-5\alpha)},\;\;\;L\equiv ln(z/z_*).
\ee
Thus the scale factor diverges and the density goes to zero providing $\alpha >1/5$. $M \rightarrow 0$ at $z=z_*$ but 
the product $MS$ tends to a constant, so the mass itself is not zero. These solutions can be regarded as asymptotically Schwarzschild. 
The condition $V_*=1$ gives a 
relationship between the constants $x_0$, $S_0$ and  $z_*$, so they are
described by two independent parameters.

\begin{figure}
\vspace*{8.0in}
\caption{This shows the form of the scale factor $S(z)$ and the velocity function $V(z)$ for the ($\alpha>1/5$) asymptotically Minkowski solutions. One family (solid lines) asymptotes to
the finite value $z_*$ with $V\rightarrow1$; the other family (broken lines) asymptotes to infinite $z$ with $V\rightarrow V_*>1$. In both cases, there are solutions
which collapse monotonically to a central singularity ($z_S$) and solutions which collapse and then bounce into an expansion phase.}
\end{figure}

\begin{figure}
\vspace*{7.0in}
\caption{This shows the form of the velocity function $V(z)$ for the
full family of spherically symmetric similarity solutions with $\alpha
=1/3$.  The exact Friedmann, Kantowski-Sachs and static solutions are
indicated by the bold lines.
Also shown (for different values of $E$) are the asymptotically Friedmann solutions and (for different values of $E$
and fixed $D$)
the asymptotically quasi-static solutions. For each value of $E$, the latter have two branches and the branches given by the broken curves all have a singularity at $z=1/D$. The negative $V$ region is occupied by the asymptotically Kankowski-Sachs solutions, although these may not be physical since the mass is negative. Solutions shown by dashed lines are irregular at the sonic point  and cannot be extended beyond there. Also shown are the two families of asymptotically flat solutions: for one of these $V$ asymptotes to $1$ at finite value of $z$; for the other $V$ asymptotes to $(2+\sqrt{13})/3$ at infinite $z$.}
\end{figure}

\section{ Applications of Spherically Symmetric Similarity Solutions}

\subsection{Gravitational collapse and naked singularities}

Until recently most counter-examples to the cosmic censorship hypothesis have been restricted to spherically symmetric spacetimes which involve shell-crossing or shell-focussing and have usually involved homothetic solutions (Eardley et al. 1986, Zannias 1991). Indeed  it has been argued that one might expect a naked singularity to {\it generally} have a horizon structure similar to that of the 
global homothetic solution (Lake 1992). The occurrence of naked singularities in spherically symmetric, perfect fluid, self-similar collapse has been studied by
Ori \& Piran (1987, 1990), Waugh \& Lake (1988, 1989), 
Lake \& Zannias (1990), Henriksen \& Patel (1991), Foglizzo \& Henriksen (1993),
Joshi and Dwivedi (1993), Joshi and Singh (1995) and Dwivedi and Joshi (1997). Their occurrence in the spherically symmetric collapse of a self-similar massless scalar field has also been studied by Brady (1995). 

Most of the early work, including the analytical studies of Eardley \& Smarr (1979) and Christodolou (1984),  focussed on self-similar dust solutions
of the Tolman-Bondi class. Ori \& Piran (1990; OP) extended this work by studying spherically symmetric homothetic solutions with pressure.
For reasonable equations of state, it might be expected that
pressure gradients would prevent the formation of shell-crossing
singularities but OP proved
the existence of a ``significant'' class of 
self-similar solutions with a globally-naked central singularity. These are the subset of asymptotically quasi-static solutions discussed in Section 2.5 with
$E_{crit}>E>E_*$. They explicitly studied the causal nature of these solutions
by analysing the equations of motion for the radial null geodesics,
thereby demonstrating that the null geodesics emerge to 
infinity. 
 
Foglizzo \& Henriksen (1993; FH) extended OP's analysis of the gravitational collapse of homothetic
perfect fluid gas spheres with $p = \alpha \mu$ to all values of $\alpha$ between 0 and 1, partially utilizing the powerful dynamical systems approach of Bogoyavlenski (1985). Their solutions can be characterized by the number of oscillations they contain in the subsonic region $n_\alpha$. As $n_\alpha$ increases, the minimum value $V_{min}$ obtained by V in the 
region $z>0$ where the singularity $z_S$ lies decreases. 
Thus there is always a first $n_\alpha$ for which $V_{min}$
falls below 1 and this value, $n_\alpha^*$ say, labels the 
threshold for the formation of massless black holes and naked singularities. FH give $n_\alpha^*=1$ for $\alpha=1/16$, $4$ for $\alpha=1/3$ and $6$ for 
$\alpha=9/16$. They surmise that $n_\alpha^*\rightarrow\infty$ as $\alpha \rightarrow 1$ but do not prove this assertion. For sufficiently small $\alpha$, $n_\alpha^*=0$ (Ori \& Piran 1990).

\subsection{Critical phenomena}

One of the most exciting developments in general relativity in recent years has been the discovery of critical phenomena in gravitational collapse. This first arose in studying the gravitational collapse of a spherically symmetric massless (minimally coupled) scalar field (Choptuik 1993). If one considers a family of imploding scalar wave packets whose strength is characterized by a continuous parameter $l$, one finds that the final outcome is either gravitational collapse for $l > l^*$ or dispersal leaving behind a regular spacetime for $l < l^*$. For $(l-l^*)/l^*$ positive and small, the final black hole mass obeys a scaling law
\be
M_{BH} = C(l-l^*)^{\gamma}
\ee
where C is a family-dependent parameter and $\gamma=0.37$ is family-independent. Initial data with $l = l^*$ evolve towards a critical solution which exhibits ``echoing''. This is a discrete self-similarity in which all dimensionless variables $\Psi$ repeat themselves on ever-decreasing spacetime scales:
\be
\Psi(t,r) = \Psi(e^{-n \Delta} t, e^{-n \Delta} r)
\ee
where n is a positive integer and $\Delta=3.44$. Near-critical initial data first evolves towards the critical solution, showing some echoing on small space scales, but then rapidly evolves away from it to either form a black hole ($l > l^*$) or disperse ($l < l^*$). 

The structure of this critical solution has been studied by Gundlach (1995). He claims that the solution is unique providing the metric is regular at the origin ($r=0$) and analytic across the past null-cone of $r=t=0$. (The null cone is also a sonic surface since the speed of sound is the speed of light for a scalar field.) The solution has a naked singularity at $r=t=0$. Gundlach also shows that the critical solution is unstable, spherically symmetric perturbations about it containing a single growing mode. A similar picture has emerged from the numerical analysis of spherically symmetric field collapse for a non-minimally coupled scalar field and for a self-interacting scalar field (Choptuik 1994).

Obtaining analytical results with a discrete self-similarity is difficult, so attempts have been made to elucidate critical phenomena by studying spherically symmetric solutions which possess continuous self-similarity. There is evidence that continuous self-similarity is a good approximation in the near-critical regime, so mathematical simplification is not the only motivation for these studies.  Spherically symmetric homothetic spacetimes containing radiation have been investigated by Evans \& Coleman (1994). They study models containing ingoing Gaussian wave packets of radiation numerically and find analogous non-linear behaviour to the scalar case. The scaling law even has the same exponent $\gamma = 0.36$, although this appears to be a coincidence since the exponent is different for other equations of state. They also obtain an exact self-similar critical solution which is qualitatitively similar to the scalar DSS critical solution; in particular, it has a curvature singularity at $t=r=0$ but is regular at $r=0$ and the sonic point. 
As in the scalar case, they find that the critical solution is an intermediate attractor: as the critical point is approached, the evolution of the fluid and gravitational field develops a self-similar region (given by the exact critical solution) near the centre of collapse. However, only a precisely critical model is described by this solution everywhere. A rather puzzling feature of the Evans-Coleman solution is claimed to be ``type 2'' since the similarity variable is $r/t^n$. However, as discussed by Carr \& Henriksen (1998), this claim is misleading. The index $n$ appears in their calculations only because their solution is asymptotically flat and this entails matching a non-self-similar region to a self-similar region across an ingoing sound-wave.

Maison (1996) has extended Evans and Coleman's work to the more general $p = \alpha \mu$ case. By considering spherically symmetric non-self-similar perturbations to the critical solution, he manages to obtain the scaling behaviour indicated by eqn (33) analytically. If $\alpha=1$, corresponding to a ``stiff'' fluid, the similarity equations are singular and so the Maison analysis cannot be straightforwardly applied to a scalar field itself. Even though a {\it homogeneous} scalar field can be described as a stiff fluid, the presence of inhomogenieties forces the equation of state to deviate from $p=\mu$. However, spherically symmetric self-similar spacetimes with a massless scalar field have been investigated by a number of authors (Brady 1995, Koike et al. 1995, Hod \& Piran 1997, Frolov 1997). The self-similar solution of Roberts (1989) is also relevant in this context. This describes the implosion of scalar radiation from past null infinity. The solution is described by a single parameter: it collapses to a black
hole when this parameter is positive, disperses to future null infinity - leaving behind Minkowski space - when it is negative, 
and exhibits a null singularity when it is zero. Although this is reminiscent of the usual critical behaviour, the critical solution is
not an intermediate attractor since nearby solutions do not evolve towards it.

It is clearly important to relate these studies of self-similar solutions to the earlier ones and to identify the critical solutions among the complete CC family. The overdense asymptotically Friedmann solutions already exhibit some of the features of the critical solution in that they are nearly static inside the sonic point and exhibit oscillations. They are also regular at the origin and at the sonic point. However, they cannot be identified with the critical solution itself since they do not contain a naked singularity at the origin. Nor can the static solution itself be so identified since it has a naked singularity at $r=0$ for all t, whereas the critical solution only has a singularity at the origin for $t=0$. 

To identify the critical solution, we need to consider the full 2-parameter family of spherically symmetric similarity solutions. If one confines attention to solutions which are analytic at the the sonic point,  then Carr et al. (1999) have shown numerically that, for all $\alpha$, the critical solution starts from a regular centre, passes through a 
sonic point and enters the spatially self-similar region (with $|V|>1$). For $0<\alpha\lesssim0.28$, the critical solution is of the
asymptotically quasi-static kind: it passes through the spatially 
self-similar region and enters a second timelike self-similar region, 
finally reaching another sonic point,
which is generally irregular. However, this does not invalidate the
solution as being the critical one, since the 
solution describing the inner collapsing region is usually matched to an 
asymptotically flat exterior geometry sufficiently far from the centre. For 
$0.28\lesssim\alpha<1$, the critical solution belongs to
the asymptotically Minkowski solutions associated with finite $z$. For the limiting case $\alpha\approx0.28$, the critical
solution is also  
asymptotically Minkowski but of the kind associated with infinite $z$.

\subsection{Self-similar voids}

Measurements of the Hubble constant $H_0$
obtained through studying Cepheid variables in galaxies in the Virgo 
and Leo clusters give values of around 80 $km\; s^{-1} Mpc^{-1}$ (Pierce et al.\  1994, Freedman et al.\  1994). 
As is well known, in the standard Big Bang model without a cosmological constant this makes it hard to reconcile the age of the Universe with the ages of the 
globular clusters (at least 12 Gyr). More recent estimates of $H_0$ using 
Cepheids yield values closer to 70 $km\; s^{-1} Mpc^{-1}$ (Freedman 1998) but
there may still be an age problem. However, it must 
be stressed that these large values of $H_0$ are all obtained 
within the relatively local distance of 100 Mpc, which is much less than
the horizon size of order 10 Gpc. Observations based on the Sunyaev-Zeldovich effect for clusters and the time delay in gravitational lensed quasars at much 
larger distances give lower values for the Hubble constant, which would be compatible with the standard cosmological age.

Several people have pointed out that the apparent discrepancy 
between the local and distant values of the Hubble constant can be 
reconciled if we live in a region of the Universe for which the local density 
is considerably less than its global value (Moffat \& Tatarski 1992 \& 1995, Nakao et al. 1995, Shi et al. 1996). This could also explain why the local density parameter (e.g. the density inferred dynamically from galaxy motions) is less than the global value required by inflation.
We will describe such  a region as a ``void'', even though it is not
completely empty. This suggestion might not seem too radical since we already 
know that the Universe contains large-scale voids (Geller \& Huchra 1989). However, to resolve the age problem, we need the local void to extend to at least 100 Mpc (so that it 
includes the Coma cluster, which is assumed to have negligible deviation 
from the Hubble flow in the Cepheid estimates of $H_{0}$) and this is much 
larger than the typical void.

In analysing this proposal, one needs to assume a particular model 
for the void. Most authors use the Tolman-Bondi solution to model the local 
region as an underdense sphere embedded in a flat Universe and then 
determine the ratio of the global and local Hubble and density parameters (Suto et al. 1995, Wu et al. 1995).  However, we have seen that density perturbations in an asymptotically Friedmann Universe may naturally
evolve to a spherically symmetric self-similar form and this has motivated Carr \& Whinnett (1999; CW) to consider the possibility that we are located at the centre of a self-similar void.

The precise form of the self-similar void depends upon 
the equation of state. After the time of decoupling at around $10^{5}$ y, 
the Universe can be treated as pressureless ``dust" ($\alpha=0$).
In this case, there is no sonic point and the 
underdense self-similar solutions can be analysed analytically as 
a special case of the Tolman-Bondi solutions (cf. Tomita 1995, 1997). CW express the various Hubble and density parameter profiles in terms of the (negative) energy parameter $E$.
Although they find that the local values of these parameters may indeed differ considerably from their global values, they also note that the self-similar dust solution is non-regular at the origin in that the circumference function is
non-zero in the limit $r\rightarrow 0$ unless $E=0$. Thus the coordinate origin is an expanding 2-sphere and must be patched onto a non-self-similar region at the centre. This produces
anomalous behaviour in the $r$-dependence of the Hubble parameter, which is in contradiction with the observations. 

Although the self-similar dust solution is not a viable model for a void in the real Universe, the similarity hypothesis is not really expected to apply in the dust situation since Carr (1993) speculates that 
one needs the effects of pressure to impose these features. It is therefore 
more natural to assume that the voids only tend to self-similar form in the 
radiation-dominated  ($\alpha=1/3$) era before decoupling.
Since the special conditions required for self-similar evolution in 
the radiation era are incompatible with self-similar evolution after 
decoupling, this suggests that one should merely use the self-similar 
radiation solution to set the initial conditions for the non-self-similar 
Tolman-Bondi evolution in the dust era. More precisely, the similarity solutions specify the form of $R$, $m$ and $E$ as functions
of $r$ along the decoupling hypersurface and the Tolman-Bondi equations then give the evolution of $R(t,r)$ for each shell of constant $r$. From this one can calculate the various Hubble parameters at any given epoch. 

CW find models that are in agreement with the observational data and clearly show a variation of Hubble parameter with distance. However, these models have a drawback. From eqn (24), the strength of the 
initial radiation perturbation is determined by the single parameter $A_{0}$,
which fixes both the density contrast and the size of the void relative to
the particle horizon. To obtain a void that is large enough to contain the Coma cluster at the present epoch, it is necessary to choose a value for $A_{0}$ which implies that the density
contrast at decoupling exceeds the mean perturbations allowed by the COBE data. In addition,
at the current epoch, the void has a local density parameter much lower than the observed value. One can select a smaller value of $A_{0}$ to
produce the required density contrast but, in this case, the void radius is too small.

\subsection {The similarity hypothesis} 

In Section 1 we emphasized that there are a variety of astrophysical and cosmological situations in which spherically symmetric solutions naturally evolve to self-similar form, leading to the ``similarity hypothesis''. Presumably a necessary (but not sufficient) condition for the validity of this hypothesis is that spherically symmetric similarity solutions (or at least some subset of them) be stable to non-self-similar  perturbations. As a first step to investigating the similarity hypothesis, Carr \& Coley (1997b) have therefore investigated the stability of the spherically symmetric similarity solutions within the more general class of spherically symmetric solutions. 

Following Cahill \& Taub (1971), we express all functions in terms of the similarity variable $z$ and the radial coordinate $r$ and regard these as the independent variables rather than $r$ and $t$. We also assume that the perturbations in $S$ and $x$ (defined in Section 2.1) can be expressed in the form
\be
S(z,r) = S_0 (z)[1+S_1 (r)] , \;\;\;  x(z,r) = x_0 (z)[1+x_1 (r)]
\ee
where a subscript 0 indicates the form of the function in the exact self-similar case and a subscript 1 indicates the fractional perturbation in that function (taken to be small; i.e. $S_1<<1$ and $x_1<<1$). The perturbations equations 
for $S_1(r)$ and $x_1(r)$ can then be expressed as 2nd order differential equation in $r$ and we test whether a particular similarity solution is stable by examining whether the perturbation terms grow or decay at large values of $r$. It might seem more natural to examine whether the solution grows or decays at large values of time, since the similarity hypothesis only requires that the general solution tends to similarity as $t \rightarrow \infty$. However, the $t$ evolution is entirely contained within the $z$ evolution, so if one wrote the perturbation equations in terms of $t$ and $z$ (instead of $r$ and $z$), one would get equivalent results.

Carr \& Coley come to the following conclusions:

$\bullet$ The asymptotically Friedmann solutions are stable providing $\alpha > -1/3$. This directly relates to the original similarity hypothesis (i.e. to the issue of whether density perturbations naturally evolve to self-similar form).  Of course, this does not prove the similarity hypothesis since they only consider {\it small} perturbations of self-similar models. Nor do they consider non-spherical perturbations.

$\bullet$ The asymptotically Kantowski-Sachs solutions are unstable for $1>\alpha>-1/3$ but stable for $-1<\alpha<-1/3$. The latter corresponds precisely to the range in which the solution may be
physically permissible. This may relate to the formation of bubbles in an inflationary scenario and hence to the stability of the inflationary phase itself.

$\bullet$ The overdense asymptotically Friedmann solutions which are non-analytic at the sonic point are unstable. This relates to the well-known ``kink'' instability: non-analytic solutions either develop shocks or are driven towards analytic ones (Ori \& Piran 1988, Whitworth \& Summers 1985).

$\bullet$ The asymptotically quasi-static solutions which contain naked singularities are unstable. This clearly relates to the 
cosmic censorship hypothesis since this states that such singularities are physically forbidden. It also relates to the instability of the critical solutions which arise in gravitational collapse calculations (Maison 1996).

\section*{ACKNOWLEDGEMENTS}

Some of the results described in this review derive from work done in collaboration with Alan Coley, Martin Goliath, Stephen Hawking, Andreas Koutras, Ulf Nilsson, Claes Uggla, Andrew Whinnett and Amos Yahil. The author thanks them for many stimulating interactions.

\end{document}